\documentclass[aps,prb,twocolumn,groupedaddress]{revtex4}
\usepackage{graphicx}

\begin{document}
\newcommand{\be}{\begin{equation}}
\newcommand{\ee}{\end{equation}}
\newcommand{\bea}{\begin{eqnarray}}
\newcommand{\eea}{\end{eqnarray}}
\newcommand{\nt}{\narrowtext}
\newcommand{\wt}{\widetext}

\title{Dirac fermions in a power-law correlated random vector potential}

\author{D. V. Khveshchenko}

\address{Department of Physics and Astronomy, University of North
Carolina, Chapel Hill, NC 27599}

\begin{abstract}
We study localization properties of two-dimensional Dirac fermions
subject to a power-law correlated random vector potential 
describing, e.g., the effect of "ripples"
in graphene. By using a variety of techniques
(low-order perturbation theory, 
self-consistent Born approximation, replicas, and supersymmetry)
we make a case for a possible complete localization of all the electronic states and 
compute the density of states.
\end{abstract}
\maketitle

\nt
Anderson localization in two spatial dimensions (2D) has traditionally been studied
in the context of doped semiconductors which are characterized by the presence of 
fermionic excitations with an extended Fermi surface.
After having been maturing steadily for almost three decades, 
the theory of 2D localization was rekindled by the advent of high-$T_c$
superconductors and, more recently, graphene, in
which systems the Fermi surface of quasiparticle excitations 
consists of isolated (nodal) Fermi points. 

Unlike in the conventional "non-relativistic" 2D electron gas (2DEG), 
in the systems of nodal fermions the single-particle density of states (DOS)
can be strongly affected by disorder.
Moreover, a greater variety of DOS behaviors and putative localization scenaria
appear to be determined not only by the symmetry of disorder, 
but its strength as well \cite{altland}. 

In undoped graphene, the low-energy Dirac-like quasiparticle excitations reside 
near two conical points
located at the momenta ${\bf K}=(4\pi/3{\sqrt 3}a,0)$ and ${\bf K}^\prime=-{\bf K}$ 
(here $a$ is the lattice spacing) in the inequivalent corners of the hexagonal Brillouin zone. 
These excitations can be conveniently described in terms of the 
(retarded) bare Green function
\be 
{\hat G}_R(\omega,{\bf p})=
[(\epsilon+i0){\hat \gamma}_0-p_\mu{\hat \gamma}_\mu]^{-1}
\ee 
where $\mu=x,y$ and the fermion velocity is $v=1$. 
The $4\times 4$ ${\hat \gamma}$-matrices 
${\hat \gamma}_{0,x,y}=({\bf 1}\otimes{\hat \sigma_3}, i{\bf 1}\otimes{\hat \sigma_2}, 
-i{\hat \sigma_3}\otimes{\hat \sigma_1})$ act in the space of Dirac bi-spinors
$\psi=(\psi_K(A),\psi_K(B),\psi_{K^\prime}(A),\psi_{K^\prime}(B))$ 
composed of the values of the 
${\bf K}({\bf K}')$-momentum components of the electron wave function on the $A$ and $B$ 
sublattices of the bipartite lattice of graphene. 

An important feature of the fermion spectrum described by Eq.(1) 
is its  "chiral" ($\epsilon\to -\epsilon$) symmetry, which, if respected
by disorder, facilitates a possible emergence of
delocalized states at $\epsilon=0$ \cite{altland}. 
Indeed, unlike in the case of the conventional 2DEG prone to 
weak localization, the conductivity of graphene
appears to be surprisingly robust, assuming 
values about three times higher than the ballistic result ($\sigma=2e^2/\pi^2\hbar$)
pertaining to the case of vanishing disorder \cite{morozov}. 

In an attempt to explain this observation, 
a number of proposals invoked the presence of an effective random magnetic field (RMF)
causing a suppression of weak antilocalization that would otherwise have occurred due
to an ordinary (chirally non-symmetric) potential disorder \cite{morozov,morpurgo}. 
In that regard, it should be noted that, being a source of elastic scattering itself, 
a RMF, too, can be solely responsible for the onset of 
localization in the conventional 2DEG, the latter 
developing in accordance with the standard unitary scenario \cite{mirlin}. 

In this Letter, we will focus on the case of 
disorder which, in the framework of the Dirac equation, 
can be described as a random (and, generally speaking, non-Abelian in the 
valley subspace) vector potential with the Gaussian variance
\bea
<{\hat A}_\mu({\bf q})\otimes{\hat A}_\nu(-{\bf q})>=
[{\bf 1}\otimes{\bf 1}w({q})\cr
+({\hat \sigma_+}\otimes{\hat \sigma_-}+{\hat \sigma_-}\otimes{\hat \sigma_+})w(2{K})](\delta_{\mu\nu}- {{q}_\mu{q}_\nu\over {\bf q^2}})
\eea
which is characterized by a parameter 
$g=na^2$ proportional to the areal density $n$ of the RMF sources ("defects")
with an internal length scale $l$. Beyond this length ($q\lesssim 1/l$) the function 
$w({q})$ decays algebraically
\be
w({q})={g/(l{q})^{2\eta}}
\ee
Although for any $\eta>0$ the parameters $g$ and $l$ can be combined together, we
choose to keep them separate. We also note that, owing to its effective nature, 
this RMF does not couple to physical fermion spin (if any).
 
With an eye on the problem of electronic transport in graphene,
we will set out to explore the general case of vector disorder with a 
variable exponent $\eta\ge 0$.
For $\eta>0$ and $l\gg a$ the inter-node scattering does get suppressed ($w(1/l)\gg w(2K)$),
and the problem can be effectively treated as a single-node one (hence, in effect, Abelian). 
We will comment on the limitations of the single-node approximation below.

The value of the exponent $\eta$ is determined by the type of 
topological structural defects, such as disclinations 
(isolated pentagon- and heptagon-rings) with $\eta=1$ and $l\sim a$, 
dislocations (pairs of adjacent pentagons and heptagons) with $\eta=0$ and $l$ 
given by the Burgers vector, etc. \cite{guinea}. 
Moreover, a mathematically similar description was also proposed for the effect of 
thermal shape fluctuations of the graphene sheet
("ripples") whose origin can be traced back to the intrinsic 
thermodynamic instability of 2D crystals. 

In recent derivations based on the theory 
of 2D elastic membranes, the effective gauge field $A_\mu$ was related
to a fluctuating local height $h$ of the graphene sheet with respect
to the substrate (the authors of Ref.\cite{antonio}
derived $A_\mu\propto (\partial_\nu^2h)^2$, 
although a much stronger dependence, $A_\mu\propto (\partial_\nu h)^2$, 
can occur as well  \cite{katsnelson}).

In what follows, we consider the more relevant version of RMF disorder 
described in Ref.\cite{katsnelson} where $g\sim 1$, $l\sim 5 nm$, 
and the exponent $\eta=0.2$ can be deduced from the previously studied asymptotic
behavior of the height fluctuations of a 2D membrane of size $L$ ($<h^2>\propto L^{1+\eta}$). 

Despite the lack of immediate physical examples, we find it instructive 
to extend our discussion to the entire range of exponents $0\le\eta<2$, for 
which the root mean square (RMS) value of the divergence-free part of the vector potential
\be
{\cal A}_\eta=[\int {d^2{\bf q}\over (2\pi)^2} w({q})]^{1/2}\propto L^{\eta-1}
\ee
(${\cal A}_1\propto \ln L/l$) increases with the system size, although the RMS
of the physical magnetic field  ${\bf B}={\bf \nabla}\times{\bf A}$ and
the corresponding energy density remain finite, ${\cal B}_\eta\propto l^{\eta-2}$.
 
It should be noted, though, that any value $\eta\neq 1$ would be in conflict 
(see Ref.\cite{dvk}) with the observed electron density dependence of the conductivity
of graphene, $\sigma\sim n_e~~~$ \cite{morozov},
if the ripples were to dominate over the other sources of elastic scattering.

While the extensively studied Abelian and non-Abelian versions of the $\eta=0$ problem
are amenable to such powerful techniques as renormalization group, 
conformal algebra, and Liouville field theory,
the case $\eta>0$ has, so far, only been studied for $\eta=1$ and 
in the ballistic regime ($\epsilon\gg 1/l$) \cite{dvk}. In order to gain a preliminary insight,
we first estimate the lowest-order correction to the (gauge-invariant) DOS
\bea
\delta\nu_\eta(\epsilon)=
Im\int {d{\bf p}{\bf dq}\over 16\pi^5}w({q})(\delta_{\mu\nu}- {{q}_\mu{q}_\nu\over {\bf q}^2})
Tr[{\hat \gamma}_\mu{\hat G}_R(\omega,{\bf p})\cr
{\hat \gamma}_0{\hat G}_R(\omega,{\bf p})
{\hat \gamma}_\nu{\hat G}_R(\omega,{\bf p+q})]
\sim g{1-\eta\over \eta}{\epsilon^{1-2\eta}\over l^{2\eta}}
\eea
which, incidentally, vanishes at $\eta=1$, while at $\eta=0$ this correction 
($\delta\nu_0(\epsilon)\sim g\epsilon\ln\epsilon$) 
is consistent with the previously obtained exact result
\be
\nu_0(\epsilon)\propto \epsilon^{2/z-1}
\ee
where $z=1+g$ for $g<2$ (weak coupling), whereas at strong coupling ($g>2$)
$z=(8g)^{1/2}-1~~~$ \cite{castillo}. 

It should be mentioned, however, that at $\eta=0$ and $l\sim a$
the original model (2,3) becomes
$SU(2)$-symmetric in the valley space ($w({0})=w(2{K})$), in which 
case the power-law DOS features a universal exponent $1/7~~~$ \cite{tsvelik}.

The first order DOS correction (5) becomes comparable to
the bare DOS ($\nu^{(0)}(\epsilon)\sim\epsilon$)
at $\epsilon\sim g^{1/2\eta}/l$, thus indicating 
the importance of higher-order corrections at still lower energies.
Routinely, a further improvement would be sought out 
in the framework of the customary self-consistent Born approximation (SCBA)
for the (gauge-non-invariant) fermion self-energy 
\be
{\hat \Sigma}_R(\epsilon,{\bf p})=\int {d{\bf q}\over (2\pi)^2}
{w(q)\over {\hat G}_R(\epsilon,{\bf p+q})^{-1}-{\hat \Sigma}_R(\epsilon,{\bf p+q})}
\ee
It can be readily seen, however, that for $\eta<1$
the SCBA solution remains finite at the nodal point, 
$\Sigma_R(0,0)=i\Gamma\sim ig^{1/2}/l$, whereas
for $\eta\ge 1$ it is determined
to the RMS of the RMF ($\Gamma\sim{{\cal A}^{1/2}_\eta}$), both diverging 
with increasing $L$.
A finite imaginary self-energy would then imply a finite 
DOS at $\epsilon=0$ (namely, $\nu_\eta(0)\sim\Gamma|\ln\Gamma a|~~$ \cite{altland}), in a 
stark contrast with Eq.(6), which calls the validity of SCBA into question.

It is well known that for $\eta=0$ the failure of SCBA stems from the fact 
that perturbative corrections represented by diagrams with crossing disorder lines 
turn out to be as important as the non-crossed ones 
(which SCBA, in effect, accounts for) \cite{tsvelik}.
However, for $\eta\ge 1$ the situation appears to be more involved, 
since a singular (small angle) scattering tends to
favor the so-called "maximally crossed" (including the nominally non-crossed)
diagrams, as can be seen, e.g., from a direct comparison 
between the crossed and non-crossed second order corrections. 

Albeit not being a justification for SCBA, this observation attests to a potentially  
singular behavior of the self-energy at small energies and/or momenta, which, in turn, 
would be indicative of a strong modification of the corresponding quantum states
by disorder that might even give rise to their complete localization.

A further evidence to that effect can be gathered from the properties of the exact
zero-energy states which can be explicitly contructed for any $\eta\ge 0$. 
In the transverse gauge representation of the random vector potential
$A_\mu=\epsilon_{\mu\nu}\partial_\nu\phi$ describing a given RMF configuration, 
the exact (unnormalized) zero-energy states read  
\be
\psi_{\pm}({\bf r})\propto ({\bf 1}\pm {\hat \gamma}_0)
\pmatrix{e^{\phi({\bf r})}
\cr
e^{-\phi({\bf r})}} 
\ee 
Among the traditional objects of interest are such quantifiers of the (normalized) 
zero-energy wave functions' statisics as the inverse participation ratios (IPR)
\be
P_n=<{\int e^{2n\phi({\bf r})}d^2{\bf r}\over 
(\int e^{2\phi({\bf r}^\prime)}d^2{\bf r}^\prime)^n}>
\ee
where the averaging over different RMF configurations is performed with the 
statistical weight 
\be
{\cal P}[\phi]=\exp[-\int {d^2{\bf q}\over (2\pi)^2}{|\phi_q|^2\over 2w({q})}]
\ee
The IPR can be computed with the use of replicas, whereby
one multiplies Eq.(9) by the denominator (normalization factor)
raised to the power $N-n$, evaluates the resulting average, 
and eventually takes the limit $N\to 0$, thus arriving at the result
\bea
P_n=\lim_{N\to 0}<\int e^{2n\phi({\bf r})+2\sum_{i=1}^{N-n}\phi({\bf r}_i)}
d^2{\bf r}\prod^{N-n}_id^2{\bf r}_i>
\cr
\sim{L^2\over l^{2n}}C_\eta(n)^{n/\eta}
\eea
where for integer values of $\eta$ 
the coefficient $C_\eta(n)$ is a logarithmic function of $L$ 
(e.g., $C_1(n)\propto\ln L/l$).

The (nearly) quadratic (for all $n$) dependence
on the system size is related to the groupping of the arguments 
($|{\bf r}_i-{\bf r}_j|\sim l$) which indicates a spontaneous symmetry 
breaking in the replica space.
Furthermore, the absence of $n$-dependence (except, for, possibly, logarithmic
factors) suggests that for $\eta>0$ the zero-energy states tend to 
(pre)localize over a characteristic length of order $l$
(possibly, up to a logarithmic factor).

This result bears a certain resemblance to that 
found in the strong coupling regime of the previously studied $\eta=0$ 
problem \cite{castillo}.
At weak coupling ($g<2$) IPR exhibit a multifractal 
spectrum $P_n\sim L^{-\Delta_n}$ governed by the anomalous dimensions 
$\Delta_n=(2-gn)(n-1)$ for $g<2$ and $n<n_c=(2/g)^{1/2}$ 
($\Delta_n=2n(1-1/n_c)^2$ for $n>n_c$), 
whereas for $g>2$ this spectrum terminates 
($\Delta_n=0$ for all integer $n$), 
thus signaling the onset of a certain "freezing" transtion \cite{horovitz}.

Another signature of this peculiar (pre)localized behavior 
can be gleaned from the correlation function of the (normalized) wave functions' amplitudes.
Using the same replica trick, the latter can be cast in the form
\bea
<\psi^2({\bf r}_1)\psi^2({\bf r}_2)>=
<{e^{2\phi({\bf r}_1)}e^{2\phi({\bf r}_2)}\over 
{(\int e^{2\phi({\bf r}^\prime)}d^2{\bf r}^\prime)^2}}>\cr
\sim \lim_{N\to 0}l^{2N-4}\exp[-{C_\eta(2)\over \eta}(N-1)({|{\bf r}_1
-{\bf r}_2|\over l})^{2\eta}]
\eea
that clearly shows an unphysical ("negative norm") behavior 
in the replica limit, which pattern is 
consistent with the replica symmetry breaking and a possible 
localization of the zero-energy states.
 
From the formal standpoint, the origin of the above 
behavior lies in the divergence of the corresponding
integral over the disorder field for any $\eta>0$ 
(at $\eta=0$ it only diverges for $g>2$).
In order to see that, we employ a 
supersymmetric formulation of the disorder averages
\bea
<X({\bar \psi},\psi)>=\int D[{\bar \psi},\psi,{\bar b},c,\phi]{\cal P}[\phi]
X({\bar \psi},\psi)\cr
\exp(i\int d^2{\bf r}[{\bar\psi}{\hat \gamma}_\mu(i\partial_\mu+A_\mu)-\epsilon)\psi+
{\bar b}{\hat \gamma}_\mu(i\partial_\mu+A_\mu)-\epsilon)c])
\eea
where $X$ stands for an arbitrary product of local fermion operators
and the path integral over bosonic ghost fields $b$ and $c$
serves to the purpose of normalizing the fermion average by the partition function before
averaging their ratio over disorder.

A further progress can be achieved with the use of the 2D bosonization technique.
Bosonization formulas \cite{ludwig}
for the relevant bilinear operators
(${\bar \psi}\psi\sim a^{-1}\cos 2\varphi$, 
${\bar\psi}{\hat \gamma}_\mu\psi=i\varepsilon_{\mu\nu}
\partial_\mu\varphi$ and ${\bar b}c\sim -a^{-1}\cos 2\theta$, 
${\bar b}{\hat \gamma}_\mu c=i\varepsilon_{\mu\nu}\partial_\mu\theta$) 
involve two bosonic ($\varphi$ and $\theta$)
and two auxiliary fermionic ($\bar \xi$ and $\xi$) fields, 
in terms of which the functional average (13) takes the form
\be
{<X({\bar \psi},\psi)>}
=\int D[\varphi,\theta,\phi,{\bar \xi},\xi]{\cal P}[\phi] 
X(e^{i\varphi})e^{-S}
\ee
and the bosonized action reads
\bea
S[\varphi,\theta,{\bar \xi},\xi,\phi]=
\int d^2{\bf r}[(\partial_\mu\varphi)^2-(\partial_\mu\theta)^2+
\partial_\mu{\bar \xi}\partial_\mu{\xi}\cr
+i(\partial_\mu\varphi
+\partial_\mu\theta)\partial_\mu\phi+{\epsilon\over a}
(\cos 2\varphi - \cos 2\theta)]
\eea
The negative sign in front of the kinetic term of the field $\theta$ 
is a salient feature of the bosonized ghost action \cite{ludwig}.

Had all the integrals over the bosonic fields been convergent, one 
could have first integrated over $\theta$,
then shifted the field $\varphi$ by $i\phi$ and, finally, integrated it out as well.
The remaining integral over $\phi$ would then 
have all the operators $\psi$ in the 
integrand replaced by $e^{\phi}$, in agreement with Eqs.(11,12). 

However, under a closer inspection one finds that the integral over
$\phi$ appears to be controled by a modified
weight ${\cal P}[\phi]\exp(\int d^2{\bf q}q^2|\phi_q|^2/(4\pi)^2)$
and, therefore, diverges for any $\eta>0$ (and $g>2$ for $\eta=0$), thereby precluding
one from the being able to readily integrate over $\varphi$ and $\theta$.

Nevertheless, the problem can be circumvented if one 
first integrates over the disorder field, thus arriving at the  
"(sub)Sine-Gordon" action with a quadratic part whose diagonal terms remain positive 
definite for the modes with momenta $q\le g^{1/2\eta}/l$
\bea
S^{(2)}[\varphi,\theta,{\bar \xi},\xi]=\int {d^2{\bf q}\over (2\pi)^2}
q^2[(1+w(q))|\varphi_q|^2\cr
+(w(q)-1)|\theta_q|^2
+|\xi_q|^2+2w(q)\varphi_q\theta_{-q}]
\eea
while its non-Gaussian part is given by the last term in Eq.(15).

Albeit not being amenable to the renormalization group treatment for any $\eta>0$,
the resulting model can still be studied with the use of the variational method.
In applications of this technique, the non-Gaussian part of the action is treated as 
a source of possible mass terms which cut off all the infrared divergencies.
We choose a variational action in the form
\bea
S_0[\varphi,\theta]=\int {d^2{\bf q}\over (2\pi)^2}
[q^2(1+w(q))+m_\varphi^2)|\varphi_q|^2\cr
+(q^2(w(q)-1)+m_\theta^2)
|\theta_q|^2+2q^2w(q)\varphi_q\theta_{-q}]
\eea
where the mass parameters 
$m_\varphi$ and $m_\theta$ satisfy the coupled equations
\be
m_{\varphi,\theta}^2={1\over L^2}<{\delta^2(S-S_0)\over \delta[\varphi,\theta]^2}>_{0}
=\pm \int D[\varphi,\theta]\cos 2[\varphi,\theta] 
e^{-S_0}
\ee
For $\eta=0$ the 
solution of Eqs.(18) reproduces the algebraically vanishing DOS (6).
The transition from weak to strong disorder and a concomitant (pre)localization
of the zero-energy states (the states at $\epsilon\neq 0 $ are localized at 
arbitrarily weak disorder) which occurs at $g=2$ can then be understood
as a proliferation of unbound vortices of the field $\varphi~~~$ \cite{horovitz}.

At all $0<\eta<2$ Eqs.(18) allow for an (approximate)
solution with the masses $m_\varphi^2=-m_\theta^2=m^2(\epsilon)$ obeying the equation
\be
\ln{m(\epsilon)\over \epsilon}={1\over 2}\int{d^2{\bf q}\over (2\pi)^2}
{q^2w(q)\over [m^2(\epsilon)+q^2]^2}
\ee 
Solving Eq.(19) with a logarithmic accuracy, one obtains the mass
\be
m(\epsilon)\sim {1\over l|\ln l\epsilon|^{1\over 2\eta}}
\ee
which reveals a chiral symmetry breaking and emergence of the chiral
order parameter $<{\bar \psi}\psi>\propto <\cos 2\varphi>$ for $\epsilon\neq 0$. 
The solution (20) applies throughout the entire range
of $g$ and $l$, thus suggesting that at $\eta>0$ 
there is no counterpart of the freezing transition 
found at $\eta=0$ \cite{castillo,horovitz}. 

The corresponding DOS is given by the expression
\be
\nu_\eta(\epsilon)={1\over \pi}Im<{\bar \psi}\psi>={\partial{m^2(\epsilon)}
\over \partial\epsilon}
\sim {1\over \eta\epsilon l^2|\ln l\epsilon|^{1+1/\eta}}
\ee
which diverges at $\epsilon\to 0$ but remains integrable for all $\eta$.
Interestingly, at $\eta=1$ the energy dependence (21) resembles
that found in the unitary limit of random potential scattering \cite{altland}
(cf. Eq.(21) with the result reported in Ref.\cite{vozmediano}).

It is worth noting that in the single-node approximation
the effect of (Abelian) potential scattering is not expected to be localizing
(which conclusion can be viewed as another manifestation of the Klein's paradox) \cite{ando}.
However, the above results indicate that, by contrast, the effect 
of a RMF on the Dirac fermions might be confining, consistent with the earlier
observations \cite{dvk,amico}.
  
On the other hand, the authors of Ref.\cite{gornyi} argued that 
in the single-node approximation the effective description of undoped graphene
can be achieved in terms of a pair of decoupled
Wess-Zumino-Witten (WZW) or (if chiral symmetry is broken)
non-linear $\sigma$ (NL$\sigma$) models, 
each of which has a topological term
driving it to a stable fixed point of either symplectic 
(if the time-reversal invariance is intact)
or unitary (as in the present case of a RMF) symmetry. 

However, the validity of the conclusions drawn in Refs.\cite{gornyi} hinges
on the possibility of performing various disorder averages
without encountering any divergencies in the underlying functional integrals.
As follows from the previous discussion, such assumptions can only be justified
in the weak-coupling regime ($g<2$) of the short-range-correlated RMF ($\eta=0$),
thus allowing for possible alternative scenarios for $\eta>0$.
In that regard, it should also be mentioned that recent numerical simulations \cite{nomura}
seem to contradict the predictions of Ref.\cite{gornyi} pertaining to the
non-chirally-symmetrical symplectic case.

The above picture also suggests that, once the 
inter-node scattering has been put back in,
the two WZW/NL$\sigma$ models become coupled and 
the topological terms cancel against each other,
thereby restoring the standard localizing behavior, which in the RMF case
would be described by the unitary NL$\sigma$-model (presumably, similar to that 
previously derived for non-relativistic spin-$1/2$ fermions 
with a gyromagnetic ratio equal two \cite{efetov}). 

Although the above scenario is believed to be rather generic,
there might be such exclusions as the $SU(2)$ (valley)-symmetric 
$\eta=0$ case of the model (2,3) where the zero-energy states remain critical
(delocalized) for all $g~~$ \cite{tsvelik}. 
Moreover, it is also possible that a non-Gaussian ($truly$ smooth) 
long-range-correlated RMF can support fermion trajectories
which follow the lines of zero field and enable semiclassical 
percolation \cite{polyakov}. 
This possibility lies well outside the domain of the Gaussian model (2,3), though.

To conclude, in this work we applied a number of techniques to gain insight
into the behavior of 2D Dirac fermions subject to a generic power-law correlated RMF.
The available evidence suggests that the zero-energy 
states become localized, while the DOS diverges at $\epsilon=0$.

The results of this work can facilitate a better 
understanding of the effects of ripples
and topological structural defects on electronic transport in graphene
and other nodal fermion systems, such as the vortex line liquid phase of high-$T_c$ cuprates.
 
This research was supported by NSF under Grant DMR-0349881.

\wt
\end{document}